\documentstyle[psfig,aps,epsf]{revtex}

\def\be{\begin{equation}}
\def\bea{\begin{eqnarray}}
\def\ee{\end{equation}}
\def\eea{\end{eqnarray}}

\def\ov{\overline}
\def\ra{\rangle}
\def\la{\langle}

\def\max{{\rm max}}
\def\thr{{\rm thr}}

\def\ea{{\it et al.}}
\def\A{{\bf A}}
\def\C{{\bf C}}
\def\a{\alpha}

\def\acc{{\rm acc}}
\def\x{{\bf x}}

\begin{document}

\begin{center}

{\large {\bf
Connectivity of neutral networks, overdispersion \\
and structural conservation in protein evolution
}}

\vspace{15truemm}

Ugo Bastolla$^1$, Markus Porto$^2$, H. Eduardo Roman$^3$ and
Michele Vendruscolo$^4$ \\

\vspace{10truemm}
$^1$ Centro de Astrobiolog\'\i a (INTA-CSIC), 28850
     Torrejon de Ardoz, Spain \\
$^2$ Max-Planck-Institut f\"ur Physik komplexer Systeme,
%               N\"othnitzer Stra{\ss}e 38,
               01187 Dresden, Germany \\
$^3$ Dipartimento di Fisica, Universit\`a di Milano,
%                Via Celoria 16,
               20133 Milano, Italy \\
$^4$ University Chemical Laboratory,
University of Cambridge, Lensfield Road, CB2 1EW, UK \\

\vspace{10truemm}

\noindent {\em Corresponding author:}\\
\vspace{3truemm}
Ugo Bastolla\\
Centro de Astrobiolog\'\i a (INTA-CSIC) \\
Tel:  +34 91 5201071 \\
Fax:  +34 91 5201621  \\
Email: ugo@chemie.fu-berlin.de  \\
Mail:  Centro de Astrobiolog\'\i a (INTA-CSIC) \\
       c.tra de Ajalvir km. 4, 28850 Torrejon de Ardoz, Spain \\

\vspace{10truemm}
\noindent{ Running title: Overdispersion and structural conservation
in protein evolution}

\noindent{{\bf Submitted to Journal of Molecular Evolution} }

\end{center}
\vspace{4truemm}

%\newpage

%\date{\today}

\centerline{\large{\bf Abstract}}
\vspace{5truemm}

Protein structures are much more conserved than sequences during evolution.
Based on this observation, we investigate the consequences
of structural conservation on protein evolution.
We study seven of the most studied protein folds, finding out that an
extended neutral network in sequence space is associated to each of them.
Within our model, neutral evolution leads to a non-Poissonian
substitution process, due to the broad distribution of connectivities in
neutral networks. The observation that the substitution process has
non-Poissonian statistics has been used against the original Kimura's neutral
theory, while our model shows that this is a generic property of neutral
evolution with structural conservation. Our model also predicts that
the substitution rate can strongly fluctuate from one branch to another
of the evolutionary tree.
The average sequence similarity within a neutral network is close to 
the threshold of randomness, as observed for families of sequences sharing
the same fold.
Nevertheless, some positions are more difficult to mutate than others.
We compare such structurally conserved positions to positions conserved
in protein evolution, suggesting that our model can be
a valuable tool to distinguish structural from
functional conservation in databases of protein families. These results
indicate that a synergy between database analysis and structurally-based
computational studies can increase our understanding of protein evolution.

\vspace{10truemm}
\noindent
\twocolumn

\section*{Introduction}

The advent of large scale genome projects is transforming the field of
molecular evolution (Koonin {\em et al.}, 2000).
%\cite{koonin00}
The molecular mechanisms of evolution are becoming increasingly amenable to
direct observation (Henikoff {\em et al.}, 1997; Gerstein, 1998: Thornton
\ea, 1999).
%\cite{henikoff97,gerstein98}
and it has become possible to study molecular evolution
not only in the context of population genetics, but also by considering
the thermodynamic stability of the biomolecules involved in evolution.
This ``structural'' approach has been pioneered by Schuster and co-workers,
with a series of studies of neutral networks of RNA secondary structures
(Schuster \ea, 1994; Huynen \ea, 1996; Fontana \& Schuster, 1998)
%\cite{RNA}
and it has been applied to proteins by several groups
(Shakhnovich \ea, 1996; Bornberg-Bauer, 1997; Bornberg-Bauer \& Chan, 1999;
Babajide \ea, 1997; Govindarajan \& Goldstein, 1997, 1998; Bussemaker \ea,
1997; Tiana \ea, 1998;
Mirny \& Shakhnovich, 1999; Bastolla \ea, 1999, 2000b; Dokholyan \&
Shakhnovich, 2001).
%\cite{BB,babadje,gold1,gold2,thirumalai,tiana,ms99,UME99,UME00,Niko}
Despite these stimulating studies, however, the structural approach has not
yet been used to investigate the classical issues in molecular evolution,
as we set out to do here.

In this work we apply to seven of the most studied protein folds
the {\it structurally constrained neutral model} (SCN)
which three of us recently introduced in the context of lattice models
%\cite{UME99,UME00}
(Bastolla {\em et al.}, 1999; Bastolla {\em et al.}, 2000b).
We compare qualitatively the substitution process obtained from the SCN
model and that observed in protein sequence evolution.
The SCN model is based on the observation that evolution
conserves protein structure much more than protein sequence
(Holm \& Sander, 1996; Rost, 1997).
%\cite{dali,rost97}
We assume that all mutations conserving the structure have the same
probability of being fixed, thus resulting in a neutral model.
The main reason to introduce structure
conservation as a working hypothesis is the experimental observation that
many mutations do not modify significantly the activity of a protein
and its thermodynamic stability, while mutations substantially improving
protein functionality are rare (Orencia \ea, 2001).
%\cite{orencia01}

The neutral theory of molecular evolution was introduced in the late
60s by Kimura (1968)
%\cite{kimura68}
and by King and Jukes (1969)
%\cite{king69}
to explain the high substitution rates observed in vertebrates
as well as the large amount of intra-specific genetic variation
observed in most species.
According to neutralism, most amino acid substitutions are fixed
in the genome of a species not because they bring a selective advantage but
due to random genetic drift acting on alleles of equivalent selective
value.
Kimura's neutral model predicts that the rate of amino acid
substitution of a given protein is approximately constant for different
species within major evolutionary groups, independent of the number of
individuals and ecology of the species. This was in agreement with
earlier observations on protein evolution which lead to postulate
a kind of ``molecular clock'' (Zuckerkandl \& Pauling, 1962).

According to Kimura's theory, the fraction of amino acids neutrally
substituted in a time $T$ is a Poissonian variable of expectation value $kT$
where $k$, the substitution rate, is different for different proteins but
does not change for different species. A subsequent study by Ohta and
Kimura (1971)
%\cite{ohta71}
measured the variance of the substitution
process acting on different species, finding that it is larger than the mean,
i.e. the process is non-Poissonian. Such a result was confirmed by the more
sophisticated analyses by Langley and Fitch (1973)
%\cite{langley73}
and Gillespie (1989).
%\cite{gillespie89}
This and other observations lead first Ohta and then Kimura to adopt,
in place of the original neutral model, a model based on slightly deleterious
mutations (Ohta, 1976),
%\cite{ohta76}
and Gillespie to reject {\it in toto} the neutral
theory favouring the hypothesis that most substitutions in protein sequences
are fixed by positive selection (Gillespie, 1991).
%\cite{gillespie91}
Takahata, however, showed that an extension of the neutral theory, the
fluctuating neutral space model (Takahata, 1987),
%\cite{takahata87}
accounts for the non-Poissonian statistics of substitutions.

One of the goals of this study is to investigate the consequences of
structural
conservation on the properties of neutral networks and on the substitution
process associated to them. We show that neutral evolution does not lead
to a Poissonian substitution process. This result complements the
fluctuating neutral space model by Takahata (1987),
%\cite{takahata87}
and suggests that arguments against the neutral theory based on the fact that
the substitution statistics is non-Poissonian (Gillespie, 1991)
%\cite{gillespie91}
are not conclusive. A deeper understanding of the mechanism of neutral
evolution will help to single out the perhaps less common but more
interesting cases of positive selection as, for instance, functional changes
and responses to changes in the environment. It can also be useful for
calibrating the molecular clocks used to reconstruct phylogenetic trees,
whose reliability is severely limited by the fluctuations of the substitution
rate (Ayala, 1997).
%\cite{Ayala}

Another interesting application of the SCN model is the possibility to
distinguish between functional and structural conservation.
By simulating neutral evolution we identify the key positions which are more
difficult to mutate. 
We identify them as structurally conserved positions, and those
positions conserved in actual evolution but not in the SCN model
as functionally conserved ones. Practically all residues whose functional
role is known belong to this class. Most positions are not
conserved in the SCN model, as similarly observed in actual evolution data. 
We identify them as neutrally evolving positions, and argue that their
preeminence is an evidence of the importance of neutral evolution. Finally,
a small number of positions appear structurally important in
the SCN model but are not significantly conserved. This could be due to a
limitation either of the SCN model or of the protein database, but it could
also be a clue of structural changes, possibly positively selected.
Other methods to identify computationally structurally important positions
have been proposed recently (Kannan \& Vishveshwara, 1999; Cecconi \ea, 2001).
%\cite{vishveshwara,fabio}
In particular, another method based on simulated evolution has
appeared in a recent preprint after this work had been completed
(Dokholyan \& Shakhnovich, 2001).
%\cite{Niko}

As most computational studies of protein evolution, the SCN model is based
on an approximate stability criterion relying on the $Z$-score
(Bowie \ea, 1991; Goldstein \ea, 1992)
%\cite{Z,Z2}
and on a folding parameter measuring the degree of correlation of the
energy landscape
(Bastolla \ea, 1999). 
%\cite{UME99}
While these parameters can not predict precisely the
thermodynamic stability of a specific sequence, our previous studies show
that they correlate with the observed stability. Thus, we expect that the
statistical properties derived from the analysis of a large number of
sequences capture real features of protein evolution.

\section*{Structurally Constrained Neutral Model}

Following Kimura, we divide mutations in two classes:
those which result in inactivating the protein, which are regarded as
lethal and can not spread in the population, and those after which the
protein remains active, which are regarded as selectively neutral.
This mutational spectrum implies that protein sequences evolve on a neutral
network, i.e. a set of sequences where the protein is active and which can
be connected through point mutations. Under this mutational spectrum,
fixation of slightly deleterious mutations can not take place, since these
are not included in the model, as well as advantageous mutations.
This is of course an important limitation of neutral models.

Kimura's neutral model assumes that the rate of appearence of neutral
mutations is constant throughout evolution. In a paper of 1977, however,
Kimura comments that rate constancy may not hold exactly (Kimura, 1977).
Our model is not based on any assumption on the neutral mutation rate.
Rather, we compute the effect of mutations on protein stability using an
effective model of protein folding (Bastolla \ea, 2000a) which provides us
with a genotype to phenotype mapping. In this respect,
the rate of occurrence of neutral mutations is an outcome of the model.
It turns out that this rate shows very broad fluctuations throughout
evolution. As we shall see below, the variance in evolutionary rates
predicted by our model is in qualitative agreement with observations of
protein evolution (Gillespie, 1991).

The model does not take into account population dynamics. This is based
on the fact that, within Kimura's model, the substitution rate is not
influenced by population size. An extension of Kimura's model to take
into account small variations in the neutral mutation rate confirmed this
result (Bastolla \& Peliti, 1991). However, population size might influence
the substitution rate if the rate of neutral mutations shows broad
fluctuations, as observed here. The explicit inclusion of population
genetics into the model would be needed to investigate this interesting
possibility.

A neutral network is defined starting from a protein sequence in the
Protein Data Bank (PDB).
The corresponding protein structure has to remain thermodynamically
stable during evolution. Thermodynamic stability is evaluated through
an effective model of protein folding. The folding parameters of the native
structure, computed through the model (see Materials and Methods), must be
above 98.5 percent of the value they have in the PDB sequence. Sequences
where this condition is met are named viable sequences. A neutral network is
a set of viable sequences which can be connected to the starting sequence
through point mutations passing on other viable sequences.
Thus sequences on a neutral network share the same protein fold and are
evolutionarily connected. For every amino acid sequence $\A$ in the neutral
network we can measure the fraction of neutral neighbors $x(\A)$, which is
the fraction of its possible point mutations which are viable.

We model protein evolution at the level of a single sequence.
During evolution, the sequence moves on the neutral network
generating an evolutionary trajectory, i.e. a list of subsequently
visited sequences belonging to the neutral network.
In the present context, the only relevant quantity is their fraction of
neutral neighbors $x(\A)\in (0,1]$. Thus an evolutionary
trajectory can be represented through a very long list
$\x=\{x_1,x_2,\cdots \}$.

The rate of occurrence of a neutral
mutation starting from sequence $\A$ is the product of a constant
mutation rate times the probability $x(\A)$ that the mutation is
viable. Thus the neutral mutation rate is not constant in the
framework of our model, and can be explicitly computed by computing
$x(\A)$ for all sequences in an evolutionary trajectory.
The statistics of the substitution process can then be obtained by coupling
the evolutionary trajectory generated as above to a Poissonian mutation
process according to the following rules:

\begin{enumerate}

\item {\bf Mutation process}: The number of mutations in a time $t$
is a Poissonian variable of average $\mu t$.

\item {\bf Acceptance process}: Given one realization of the
evolutionary trajectory and a number of mutations $k$, the conditional
probability that $n$ of them are accepted is the product of $n+1$
geometric distributions of parameters $1-x_i$:

\be
P_{\acc}\left(n\mid k\right)=
\left( \prod_{i=1}^{n}x_i \right)
\sum_{\{m_j\}}\prod_{j=1}^{n+1}(1-x_j)^{m_j}\; ,
\label{acc}
\ee
where the $\{m_j\}$ are all integer numbers between zero and $k-n$
satisfying $\sum_{j=1}^{n+1}m_j=k-n$. In other words, the probability that a
mutation is accepted is $x_1=x(\A_1)$ as long as the protein sequence is
$\A_1$, $x_2=x(\A_2)$ as long as the sequence is $\A_2$, and so on.

\end{enumerate}
Two kinds of random variables must be distinguished. We indicate by angular
brackets the average over mutation and acceptance process for a given
realization of the evolutionary trajectory and by an overline the average
over evolutionary trajectories. The variance of the substitution process can
be decomposed in two components:

\bea
V(S_t) & = & V_{\mu}(S_t)+V_x(S_t) \label{variance}\\
& = &
\left(\ov{\la S_t^2\ra} -\ov{\la S_t\ra^2} \right)+
\left(\ov{\la S_t\ra^2} -\ov{\la S_t\ra}^2 \right)\; .
\nonumber
\eea

The first term, $V_\mu$, is the variance of the mutation and
acceptance process, averaged over evolutionary trajectories.
The second term, $V_x$, is the variance of the substitution rate with
respect to different evolutionary trajectories. This term, which is not
present in the standard neutral model, explains why the variance of the
number of substitution is typically larger than its mean value, contrasting
with a Poissonian process.

If all sequences have the same fraction of neutral neighbors $x(\A)\equiv x$,
the number of substitutions in a branch of length $T$ is Poissonian with
mean $\mu T x$ and the substitution rate is equal to $\mu x$ as in Kimura's
model. If $V_x$ is not zero, the substitution distribution is more
complicated and has to be computed numerically using the
evolutionary trajectories simulated.

\section*{Numerical results}

\subsection*{Folding of random sequences}

As a preliminary analysis, we measured the distribution of the folding
parameters $\alpha$ and $-Z'$ (see Materials and methods) of the native
structures considered in this work for random sequences of the same length
of the corresponding PDB sequences. On over 20,000 attempts, we always found
folding parameters much lower than for PDB sequences.
The only exception was the smallest protein, the 53-residues rubredoxin, for
which a single random sequence had one of the stability parameters
comparable to that of the PDB sequence, even if still smaller than it.
This result is consistent with the work of Keefe and Szostak who were
recently able to select ATP binding proteins from a random library of more
than $10^4$ sequences (Keefe \& Szostak, 2001).
%\cite{keefe01}
We note that it is possible to evaluate the size of the
neutral network from the joint distribution of $\alpha$ and $Z'$ ,
but to this end better statistics are needed than those obtained here.

\subsection*{Connectivity of neutral networks}

\begin{figure}
 \centerline{
    \psfig{file=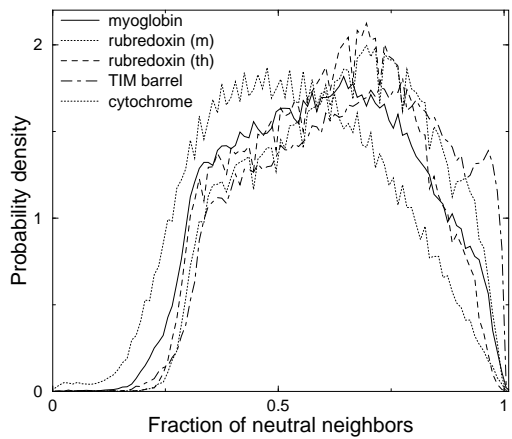,height=6.0cm,angle=0}
 }
\caption{Distribution of the fraction of neutral neighbors for four
protein folds.}
\label{distr}
\end{figure}

For sequences $\A$ belonging to the neutral network, the fraction
of neutral neighbors $x(\A)$ counts the fraction of all possible point
mutations of $\A$ which still fall into the neutral network. We measured
this quantity for at least 20,000 sequences for each fold, finding
that it has a broad distribution (see Fig. \ref{distr}). The
shape of the distribution is qualitatively similar for all of the studied
proteins, but in the case of cytochrome c, the distribution is shifted to
lower connectivities. Results for all seven folds are summarized in Table 1.

The connectivity landscape $x(\A)$ is locally correlated for distances
in sequence space of the order of at least ten substitutions.
The correlation function
$C(t)=\left(\left\langle x(\A_{t_0})x(\A_{t_0+t})\right\rangle
-\left\la x(\A) \right\ra^2\right)/\sigma_x^2$
between connectivities of two sequences at distance of $t$ steps
decays similarly for all proteins and can be fitted to a stretched
exponential $C(t)\approx \exp\left(-(t/\tau)^\eta\right)$, with exponents
$\eta$ ranging from $0.60$ to $0.66$ and correlation lengths $\tau$ ranging
from $1.8$ to $2.8$. Thus correlations decay to one tenth after
about ten substitutions (data not shown).

\subsection*{Substitution process}

\begin{figure}
 \centerline{
      \psfig{file=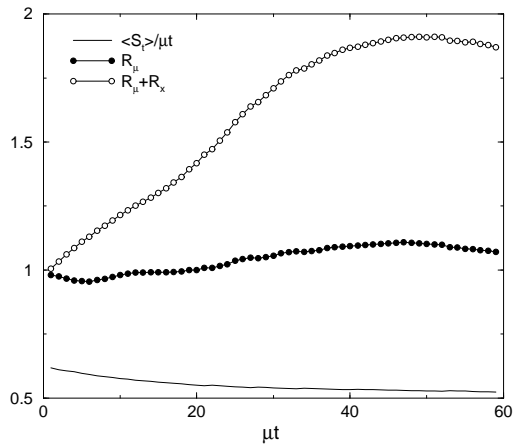,height=6.0cm,angle=0}}
\caption{Moments of the substitution process for myoglobin {\it versus}
mutational time $\mu t$. Solid line: average substitution rate divided by
$\mu$,
$\overline{\langle S_t\rangle}/\mu t$. Black circles: mutational variance
$\overline{\langle S_t^2\rangle-\langle S_t\rangle^2}$
divided by the mean ($R_\mu$).
White circles: total dispersion index, $R_\mu+R_x$,
where $R_x$ is the trajectory variance
$\overline{\langle S_t\rangle^2}-\overline{\langle S_t\rangle}^2$
divided by $\overline{\langle S_t\rangle}$.}
\label{myog_subst}
\end{figure}

The broadness of the connectivity distribution directly implies that
the substitutions process fluctuates more than a Poissonian process, i.e.
it is overdispersed. We computed average and variance of the substitution
process numerically, using the evolutionary trajectories generated in our
simulations. Results for myoglobin are shown in Fig.~\ref{myog_subst}.
Notice that the substitution rate $\overline{\langle S_t\rangle}/ t$ is
roughly
constant in time, and the total dispersion index
$R(t)=\overline{V(S_t)}/\overline{\langle S_t\rangle}$
takes values between 1.0 at small $t$ and 1.9 at large $t$, consistent with
the value $R=1.7$ estimated by Kimura for myoglobin.

\begin{figure}
 \centerline{
      \psfig{file=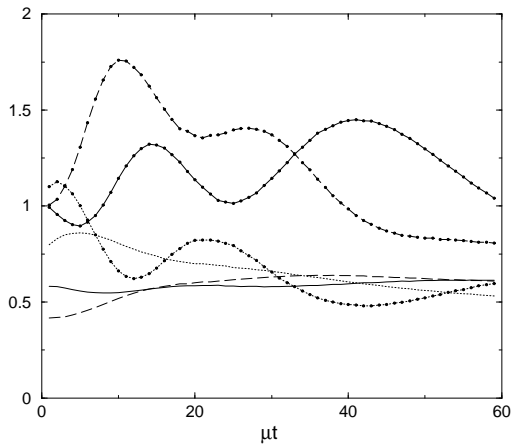,height=6.0cm,angle=0}}
\caption{Substitution process for three realizations of the evolutionary
trajectories of myoglobin. Black circles represent
variances, no symbols represent mean values.}
\label{myog_subst2}
\end{figure}

Due to local correlations in sequence space,
different evolutionary trajectories $x(\A_1)\cdots x(\A_n)$,
representing different populations, give different mean and variance
of the substitution process over short time scales.
This phenomenon produces new lineage effects, i.e. apparently varying
substitution rates in different branches of the phylogenetic tree.
To illustrate them,
we show in Fig. \ref{myog_subst2} the mean $\langle S_t(\{x\})\rangle$ and
the variance $V(S_t,\{x\})$ for three realizations of the evolutionary
trajectory $\{x\}$. Such an effect could overshadow the generation time
effect for replacement substitutions
(Britten, 1986; Li \ea, 1987; Gillespie, 1991).
%\cite{gillespie91,li87,britten86}
It could also been responsible for the wide fluctuations in the
substitution rate for different lineages observed by Ayala (1997).
%\cite{Ayala}

\subsection*{Sequence similarity}

\begin{figure}
 \centerline{
      \psfig{file=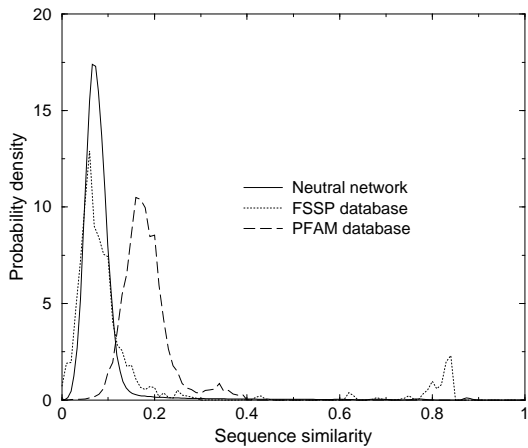,height=6.0cm,angle=0}
}
\caption{Sequence similarity distribution for the myoglobin fold.}
\label{myog_over}
\end{figure}

One may ask whether sequences sharing the same fold must have a high
level of similarity. To investigate this question, we measured the
distribution of sequence similarity for sequences in the neutral network
obtained through simulation of our model as well as for homologous
sequences in public databases. Similarity between two aligned sequences of
the same length is defined as the fraction of positions where the same
amino acid appears.
Results are showed in Fig.~\ref{myog_over} for the globin fold. Similar
results have been obtained for all other folds.

Sequences in the neutral network
(solid line) have an average similarity only slightly larger than random
sequences, for which a Gaussian-like distribution of average value $1/20$
is expected. This confirms a previous finding of three of us for neutral
networks of lattice structures (Bastolla \ea, 1999).
%\cite{UME99}
As previously observed by Rost (1997), the same result holds
also for sequences in the FSSP family of sequences sharing the same
structure (Holm \& Sander, 1996), which are showed as dotted line.
%\cite{dali}
Low similarity for sequences with the same fold has also been found in a
recent computational study based on sequence optimization for native protein
structures (Dokholyan \& Shakhnovich, 2001).
%\cite{Niko}.
The dashed line shows the similarity distribution for sequences in the
PFAM database of similar sequences (Bateman \ea, 2000).
The PFAM family has on the average a much larger
similarity. This is in part due to the fact that in this case sequence
similarity must be large enough for the homology to be detected and in part
to the fact that proteins in a PFAM family are subject to stronger functional
conservation than proteins in the FSSP family. For the globin family, which
has been intensively studied, even the PFAM similarity is very low.

\subsection*{Residue conservation}
The SCN model identifies structurally conserved residues. In this respect,
the present results can be compared to the model of Shakhnovich and
coworkers (Shakhnovich \ea, 1996; Dokholyan \& Shakhnovich, 2001)
%\cite{shak96,Niko}
based on sequence optimization (Shakhnovich \& Gutin, 1993; Shakhnovich,
1994),
%\cite{SGinv}
and to the bioinformatic studies of Ptitsyn's group (Ptitsyn, 1998;
Ptitsyn \& Ting, 1999).
%\cite{ptitsyn98,ptitsyn99}

We evaluate the conservation of each position measuring its rigidity
(see Materials and Methods).
This was done for each fold using three different sets of sequences:
(i) Sequences obtained from simulations of our neutral model; (ii)
Homologous sequences in the PFAM database (Bateman \ea, 2000); (iii)
Sequences with the same structure in the FSSP database (Holm \& Sander, 1996).
%\cite{dali}
The PFAM family often contains orthologous proteins
performing the same function while in the FSSP family different functions
may be present and function conservation plays a less important role.
Nevertheless, there is usually a good correlation between the rigidity of
a given position evaluated through the PFAM and FSSP databases.
An exception is the TIM barrel family, one of the most common folds, used to
perform different functions, each approximately associated to a different
PFAM family. In this case, the two sets of rigidities show no correlation.
For this fold, using a bioinformatic analysis, Mirny and Shakhnovich (1999)
found evidence of functional conservation (the same functional positions
tend to be conserved in all functional families, although with different
residues in each functional family) but could not find evidence of
structural conservation.

Before turning to the analysis of conserved positions, we observe that there
are three reasons why sequence databases
may tend to overestimate structurally based conservation. The first one is
the small size of databases. The second one is the fact that many sequences
are evolutionarily related: databases usually provide biased samples of the
tree of life. To reduce these effect, we limit our analysis to
sequences that do not have similarity larger than a threshold $q_{\rm thr}$
which we choose equal to 0.85 in order not to reduce too much database
size, and try to estimate the maximal conservation that one would observe
with a database of similar size and correlations, in the null hypothesis
that all positions are equivalent. The third reason is that many residues
are conserved on functional grounds, sometimes even in the FSSP database,
and it may be difficult to distinguish them from structurally conserved
residues.

Conservation in the neutral network only expresses structural conservation,
thus the comparison between rigidities predicted by the SCN model and
observed in evolution may allow to single out functionally conserved
positions or positions involved in interactions with cofactors, which are not
represented in our model. We tested this for two well studied
protein families: the globin
family and the cytochrome c family. In both cases structurally conserved
positions identified by the SCN model coincide with those identified in
previous bioinformatic studies as part of the folding nucleus
(Ptitsyn, 1998; Ptitsyn \& Ting, 1999),
%\cite{ptitsyn98,ptitsyn99}
and additional structurally conserved positions
are found. For other protein families less is known about
functional residues, but the few ones which are identified in the SwissProt
file are recognized as such by the SCN model.
In our analysis of the globin family, positions in contact with the
heme are not regarded as structural, even if the heme plays also an
important stabilizing role, since interactions between amino acids and
cofactors are not considered in the model, and they are much more specific
than interactions between amino acids.

\begin{figure}
 \centerline{
      \psfig{file=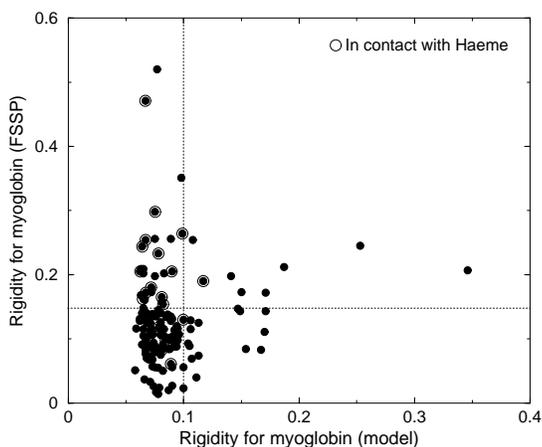,height=6.0cm,angle=0}
}
\caption{Rigidity in the FSSP family {\it versus} rigidity in the neutral
network for the myoglobin fold.}
\label{myog_rigid}
\end{figure}

In Fig.~\ref{myog_rigid} we compare the rigidities obtained from our model
to those measured in the FSSP family for the myoglobin fold. Each point
represents a position on the native structure. 
The dotted lines in the figure are rough estimates of
the maximal rigidity expected in a random situation, i.e. all equivalent
residues and same distribution of similarity as in the set examined. Only
residues more rigid than that are considered significantly conserved.
Many of the most conserved residues are in contact with the heme group
(large circles). A notable exception is Pro37\footnote{
Residues are labeled in the order in which they are listed in the PDB file
of the structure 1a6g}, which is strongly conserved and not in
contact with the Heme group. Although the conservation of this residue has
not been fully explained so far, Ptitsyn and Ting report that it may be
due to functional reasons (Ptitsyn \& Ting, 1999).
%\cite{ptitsyn99}
The three positions most conserved according to the SCN model coincide
with structural positions identified in the bioinformatic analysis by
Ptitsyn and Ting.
They are, in order of rigidity, Leu115, Trp14, Met131. Val10 is rather
conserved both in our model and in the bioinformatic study. The remaining
two positions identified by Ptitsyn, Ile111 and Leu135, are not among the
most conserved in our model, although they are above the average.  
In addition, there are eight more positions significantly conserved in the
SCN, whose evolutionary conservation is somewhat lower (but above the
average in all cases except His119). They
are: Val13, Val17, His24, Leu69, Leu76, His119, Phe123, Ala134.
Interestingly, structurally conserved residues form a cluster, so that
it has been proposed that they play the role of ``folding nuclei''
(Shakhnovich \ea, 1996; Ptitsyn 1998; Mirny \& Shakhnovich, 2001).
%\cite{shak96,ptitsyn99,ms01}
A similar situation applies also for the
case of cytochrome c: two of the positions identified in (Ptitsyn, 1998)
are the most conserved in our model (Phe7, Leu74), another one is
significantly conserved (Trp77) and the fourth one is not present in the
structure we choose as reference (PDB code 451c). Moreover, there are
three positions significantly conserved in our model and in the FSSP
alignment (Tyr27, Ile48, Val66) and one conserved in our model but not
in the alignment (Gly36).

\begin{figure}
 \centerline{
      \psfig{file=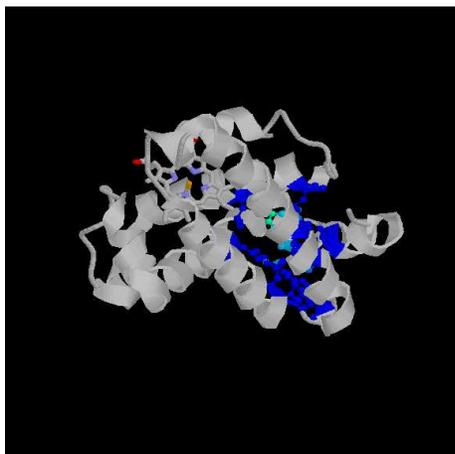,height=6.0cm,angle=0}
}
\caption{Structure of myoglobin with the Heme group. The structurally most
conserved residues are represented in color together with their side
chains. These are: Leu115, Trp14, Met131, Val10,
Val13, Val17, His24, Leu69, Leu76, His119, Phe123, Ala134.
The PDB code of the structure is 1a6g.
%the figure is made with rasmol.
The colour code represents temperature increasing from blue to red.
}
\label{myog_fig}
\end{figure}

\begin{figure}
\centerline{
      \psfig{file=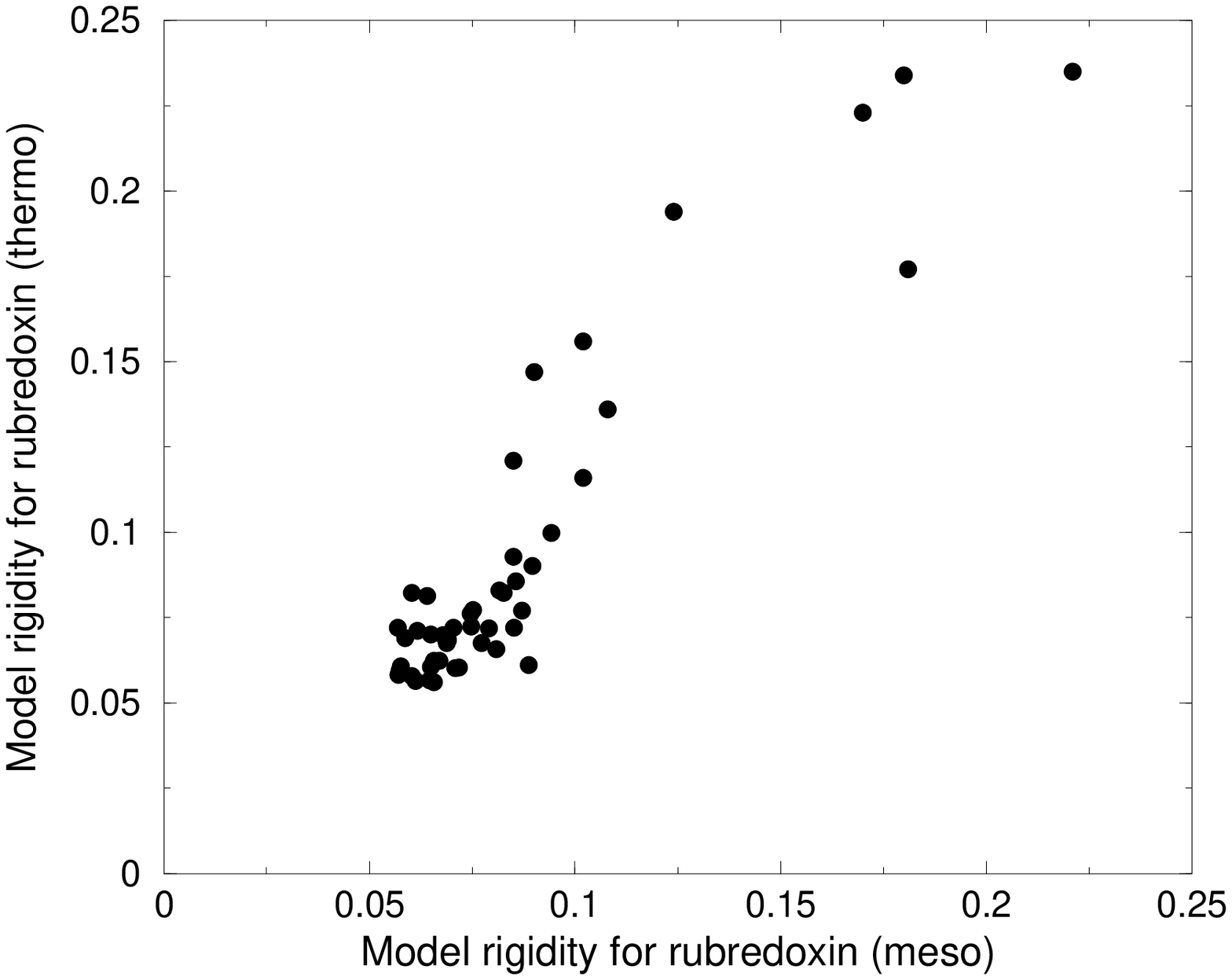,height=6.0cm,angle=0}
}
\centerline{
      \psfig{file=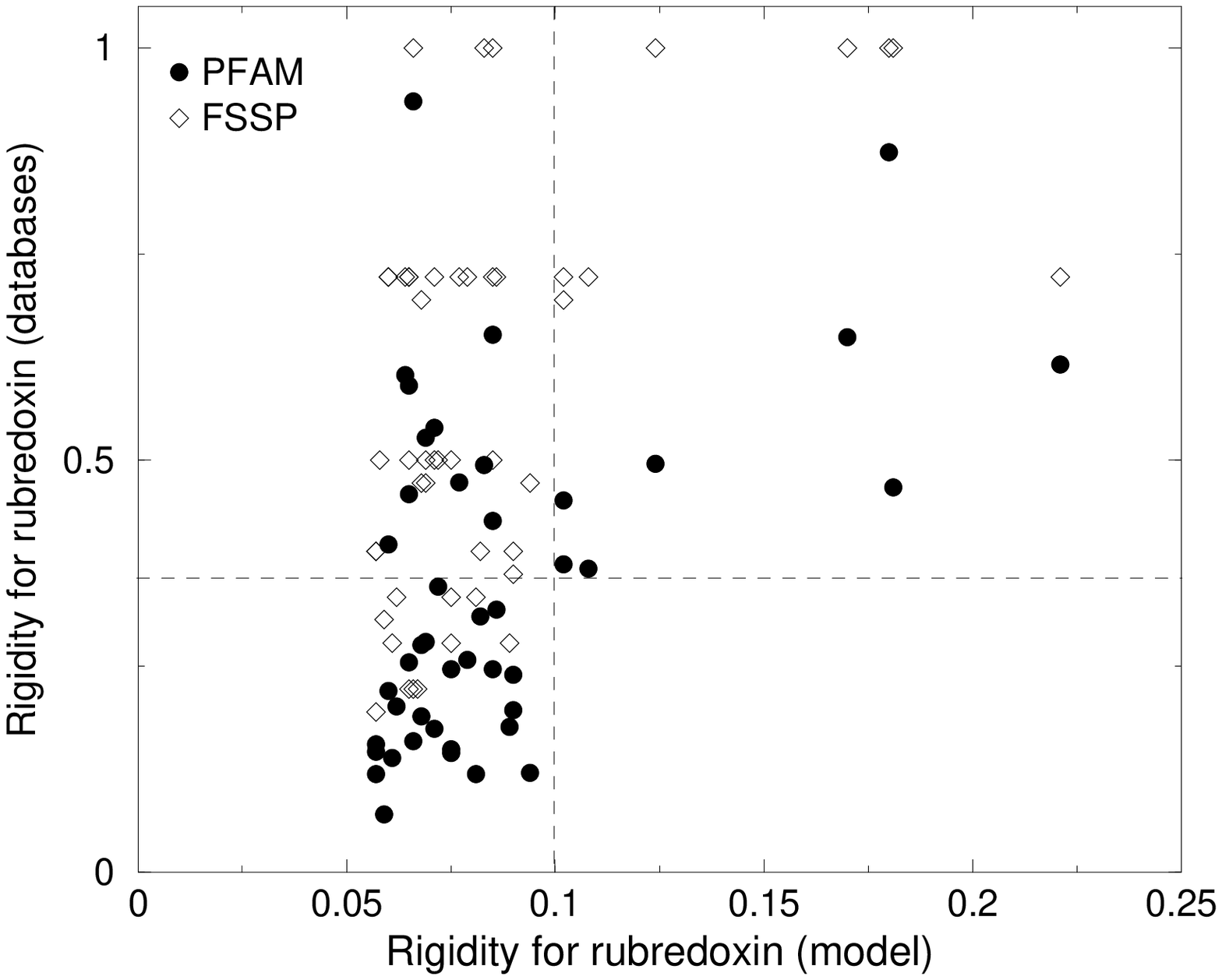,height=6.0cm,angle=0}
}
\caption{Comparison of rigidities between the two simulated neutral networks
of mesophylic and thermophylic rubredoxin (top) and between simulated and
observed rigidities for mesophylic rubredoxin (bottom).
Full circles refer to the PFAM family, open diamonds to the FSSP family.}
\label{rub}
\end{figure}

We show a similar plot also for rubredoxin, a small bacterial protein
involved in electron transport. In this case,
we studied two homologous proteins, one from a mesophylic and one from a
thermophylic bacterium. Their sequences have 57\% similarity and belong to
the same PFAM and FSSP classes. Although the structures are rather similar,
the stability of the thermophylic protein, as measured by the $Z'$ and $\a$
parameters, is higher than the stability of the mesophylic protein, as
it should be. This result supports our choice of the stability criteria.
We compare the rigidities obtained from the SCN model for the two structures
in Fig.~\ref{rub} (upper panel). There is a remarkable correlation, despite
the fact that results are obtained from independent evolutionary runs with
different selection parameters. In Fig.~\ref{rub} (lower panel)  we compare
the rigidity observed in the SCN model (for mesophylic rubredoxin) with the
rigidities observed in the PFAM and FSSP databases.

\begin{figure}
 \centerline{
      \psfig{file=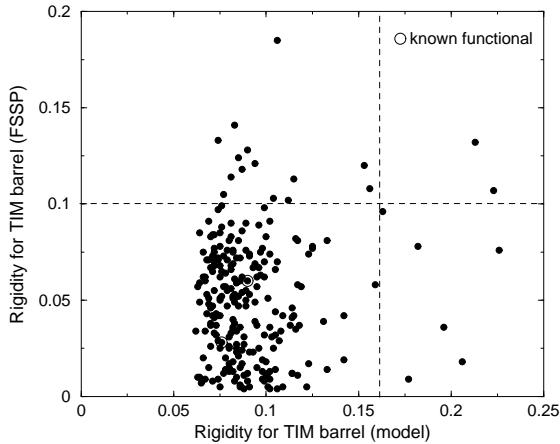,height=6.0cm,angle=0}}
\caption{Rigidity in the FSSP database versus rigidity in the neutral network
for the TIM barrel fold. The circles identify two positions of known
functional role for one of the enzymes of the TIM barrel family.}
\label{tim}
\end{figure}

Finally, we show in Fig.~\ref{tim} a scatter plot of rigidities for the
protein showing the worst correlation between predicted and observed
rigidities: the TIM barrel, one of the most common folds, used
for several enzymatic functions. The one that we studied is a triose
phosphate isomerase functioning in the glycolysis. In this case, there is
also no observable correlation between rigidities in the PFAM and FSSP
databases, and rigidities in the FSSP class are very low, in particular
because several residues are deleted in many sequences.

%Figure \ref{ubi_fig} shows the most conserved residues in the neutral network
%of ubiquitin.
Unlike for other folds, in the case of ubiquitin the
structurally conserved positions are distributed along the main axis of the
protein. There is also a conserved polar position on one loop.
For lysozyme as well the most conserved residues form a non compact core.
In all other cases the structurally conserved residues form a hydrophobic
cluster which is rather compact. There is some correlation between
conservation and burial in the interior of the protein as measured by the
number of contacts, but burial alone does not explain all of the conservation.
We notice that also in our previous lattice simulation the most conserved
positions were those in the interior of the lattice structure
(Bastolla \ea, 1999).
%\cite{UME99}
%\begin{figure}
% \centerline{
%      \psfig{file=ubi2.ps,height=6.0cm,angle=0}
%}
%\caption{Structurally conserved residues for the ubiquitin fold:
%W41, Ile45, Ile109, Leu113, Leu120, Gln126, Tyr144. Notice they are all
%aligned along the proteins's axis, except Gln126 which protrudes from
%one loop. The PDB code of the structure is 1u9a.
%%the figure is made with rasmol.
%The colour code represents temperature, increasing from blue to red.
%}
%\label{ubi_fig}
%\end{figure}

\section*{Discussion}

In this work we studied a neutral model of protein evolution based on
structure
conservation. For all of the examined folds, local connectivities of neutral
networks are broadly distributed. This result implies that even in neutral
evolution the number of substitutions fluctuates more than a Poissonian
variable, i.e. it is overdispersed.
Therefore overdispersion can not by itself be used as a test for positive
selection, as argued for instance by Gillespie (1991).
%\cite{gillespie91}
Our results, nevertheless, show that the dispersion index of the
SCN substitution process is typically small, rarely overcoming four.
Thus, proteins with extremely high dispersion index, as some of those studied
by Gillespie (1989)
%\cite{gillespie89}
or Ayala (1997)
%\cite{Ayala}
are not likely to have
evolved in this way. The average substitution rate is almost constant in
time, but it may vary considerably for different evolutionary trajectories,
corresponding to different branches of the phylogenetic tree. This fact
should be taken into account when studying lineage effects such as the
generation time effect
(Britten, 1986; Li \ea, 1987; Gillespie, 1991).
%\cite{gillespie91,li87,britten86}.

By simulating neutral evolution, we identified structurally conserved
positions and compared them to evolutionarily conserved positions in known
protein families. The comparison is favorable for myoglobin, cytochrome c,
lysozyme, ribonuclease and rubredoxin, while for ubiquitin and the TIM barrel
correlation between predicted and observed conservation is almost absent. We
note, however, that the TIM barrel shows very little structural conservation,
and the small size of the ubiquitin family makes the comparison not
conclusive.
The plots comparing conservation in simulated evolution (on the abscissa)
to conservation in real evolution (on the ordinate) can be divided in four
parts. In the upper left quadrant there are positions not conserved in the
SCN model but conserved in evolution. We suggest that most of them are
conserved for functional reasons or because of interactions with cofactors,
which are not taken into account in our protein model.
Positions of known functional importance belong to this class, but not enough
is known on protein function to prove our interpretation in all cases.
In the upper right quadrant there are positions conserved both in the SCN
model and in the databases, whose conservation is likely to have a structural
ground. For this small subset the rigidities that we predict are correlated
to the observed ones.
Interestingly, those positions form spatial clusters which have been
identified with folding nuclei
(Shakhnovich \ea, 1996; Ptitsyn, 1998; Mirny \& Shakhnovich, 2001).
%\cite{shak96,ptitsyn98,ms01}
Although we can not discuss such interpretation, since our evolutionary
algorithm does not take into account folding kinetics, it is to be expected
that positions important for stability also play an important kinetic role.
In the bottom left quadrant there are positions not
conserved neither in the SCN model nor in the databases. These positions are
likely to be the main actors in neutral evolution. Last, in the bottom
right quadrant there are few positions conserved according to the SCN model
which do not appear to be evolutionarily conserved. Barring artifacts due to
the SCN model, we should consider the possibility of conservation with low
rigidity (but typically much higher than random).
In order to verify whether this is the case, we need larger and less
correlated protein classes. Another possibility is that these positions are
frequently substituted because they can produce structural changes, possibly
positively selected. Although this possibility is rather speculative,
it would be interesting to investigate it in more detail.

Our results are based on an approximate stability criterion relying on the
$Z$-score (Bowie \ea, 1991; Goldstein \ea, 1992)
%\cite{Z,Z2}
and on a parameter measuring the degree of correlation of the energy
landscape (Bastolla \ea, 1999).
%\cite{UME99}
While such criterion may
not be suitable for the quantitative prediction of the thermodynamic
stability of particular proteins, we believe that the
statistical properties of the SCN model reflect those of actual protein
evolution. This confidence has several grounds. First, in the study of a
lattice model, three of us have previously applied a rigorous criterion of
stability, and compared it to a criterion obtained from the
$Z$-score (Bastolla \ea, 2000b).
%\cite{UME00}
Although the two criteria give different responses
for specific sequences, it is possible to choose a threshold such that most
sequences selected with the $Z$-score criterion are also selected with the
rigorous criterion. Second, the present results are robust with respect to
changes in the selection thresholds and stability criteria. Third, we
tested our stability parameters on a large number of sequences obtained
from mutations of a TIM barrel enzyme, whose phenotypic effect has been
experimentally measured in a recent paper (Silverman \ea, 2001).
%\cite{silverman01}
We found that, even if our criterion can not predict the effect of individual
mutations, the latter is correlated to the $\a$ parameter with correlation
coefficient 0.4.

The present results can be used in the rational approach to directed
evolution of biocatalysts (Altamirano \ea, 2000; Voigt \ea, 2001)
%\cite{altamirano00,voigt01}
since we identify sites that are more tolerant to mutations and therefore
likely targets for evolutionary improvement. This is a remarkable possibility,
since it indicates how results based on the assumption of neutral evolution
can be used to search for positive substitutions.

\section*{Materials and Methods}

\subsection*{Protein model}
We represent a protein structure by its contact matrix $C_{ij}$,
where $C_{ij}=1$ if residues $i$ and $j$ are in contact and $C_{ij}=0$
otherwise. Two residues are considered in contact if any two of their heavy
atoms are closer than 4.5 \AA. The effective free energy associated to a
sequence of amino acids $\A$ in the configuration $\C$ is approximated
as a sum of pairwise contact interactions,
\be E(\A,\C)=\sum_{i<j} C_{ij}U(A_i,A_j),
\ee
where $A_i$ labels one of the twenty amino acid types and $U(a,b)$ is a
$20\times 20$ symmetric interaction matrix. Here we use the matrix derived by
Bastolla {\it et al.} (2000a),
%\cite{opt1}
which describes accurately the thermodynamic stability of a large set of
monomeric proteins (Bastolla \ea, 2001).
%\cite{opt2}

Three remarks are needed: First, the effective energy parameters implicitly
take into account the effect of the solvent and depend on temperature. They
express free energy rather than energy. Second, the effective energy of a
structure is defined with respect to a completely extended reference
structure where no contacts are formed and which sets the zero of the
energy scale. Third, one can derive from the database not the parameters
$U(a,b)$ themselves but the dimensionless quantities $U(a,b)/k_B T$. It is
thus important to use dimensionless parameters to evaluate the stability
of the protein model.

\subsection*{Candidate structures}

We generate candidate structures for a protein sequence of $N$ residues
by generating all possible gapless alignments of the sequence with
structures in the Protein Data Bank. This procedure is called {\it threading}.
In this way, we typically generate several hundreds of thousands of
protein-like structures per sequence. In the present context,
threading is directly used to produce the contact maps of the candidate
structures. In order to speed up the computations, we use a non
redundant subset of the PDB excluding proteins with homologous sequences,
selected by Hobohm \& Sander (1994).

\subsection*{The folding parameter $\a$}

For a given sequence $\A$, the energy landscape is well correlated if all
configurations of low energy are very similar to the configuration of minimal
effective energy, $C^*(\A)$. Structure
similarity is measured by the overlap $q(\C,\C^*)$, counting the number of
contacts that two structures have in common and normalizing it through
the maximal number of contacts, so that $q$ is comprised between zero and one.
In a well correlated energy landscape, the inequality holds

\be
{E(\A,\C)-E(\A,\C^*)\over |E(\A,\C^*)|}\geq \a(\A)
\left(1-q(\C,\C^*)\right)\, , \label{alpha}
\ee
stating that the energy gap between each alternative structure $\C$ and the
ground state $\C^*$, measured in units of the ground state energy, is
larger than a quantity $\a(\A)$ times the structural distance $1-q(\C,\C^*)$.
The dimensionless quantity $\a(\A)$, which is the largest quantity for which
the above inequality holds, can be used to evaluate the folding
properties of sequence $\A$. For random sequences, many different
configurations have quite similar energy and $\a(\A)\approx 0$. In this case
the energy landscape is rugged, the folding kinetics is very slow and the
thermodynamic stability with respect to variations in the solvent is very
low. In contrast, computer simulations of well designed sequences have
shown that, when $\a(\A)$ is finite, the folding kinetics is fast and the
stability with respect to changes in the energy parameters as well as
mutations in the sequence is very high.

Our algorithm computes the parameter $\a(\A)$ for a fixed target
configuration $\C^*$ and a large number of sequences $\A$.
We thus indicate this parameter as $\a(\A,\C^*)$, since we do not know
a priori that $\C^*$ has lowest energy. Notice however that, if
$\a(\A,\C^*)$ is positive, all alternative structures
have higher energy than $\C^*$.
We impose that $\a(\A,\C^*)$ is larger than a positive threshold $\a_{\thr}$
for sequences $\A$ belonging to the neutral network.

\subsection*{$Z$-score}

The Z-score $Z(\A,\C^*)$ (Bowie \ea, 1991; Goldstein \ea, 1992)
%\cite{Z,Z2}
is a measure of the compatibility between a sequence $\A$ and a structure
$\C^*$, widely used in structure prediction. It depends on an effective
energy function, and measures the difference between the energy of sequence
$\A$ in configuration $\C^*$ and its average energy in a set of alternative
configurations, $\{ \C\}$, in units of the standard
deviation of the energy:

\be
Z(\A,\C^*)=\frac{E(\A,\C^*)-\la E(\A,\C)\ra }
{\sqrt{\la E(\A,\C)^2\ra-\la E(\A,\C)\ra^2}} \: .
\ee

When sequence $\A$ folds in structure $\C^*$ their corresponding
Z score is very negative.

Given the above definition, one has still to specify
how to select alternative structures. A possibility, often used for lattice
models (Mirny \& Shakhnovich, 1996)
%\cite{MS}
is to assume that alternative structures are maximally
compact, randomly chosen structures, whose average energy can be estimated
as $\la E(\A,\C)\ra_{\C}=N c_{\max} \la e(\A)\ra$.
Here, $Nc_{\max}$ is the maximal number of contacts of candidate
structures and $\la e(\A)\ra$ is the average energy of a contact, averaged
over all possible contacts formed by sequence $\A$. This leads to introduce
the parameter

\be
Z'={E(\A,\C^*)/N c_{\max}-\la e(\A)\ra\over
\sqrt{\la e^2(\A)\ra-\la e(\A)\ra^2}}\, .
\ee

The use of $Z'$ has two main advantages: First, it makes the value of the
$Z$-score much less sensitive to chain length $N$ and to the particular set
of alternative structures used; second, the evaluation of $Z'$ is much faster
than that of the $Z$-score. This is necessary in order to explore efficiently
sequence space.

\subsection*{Sampling the neutral network}

Our algorithm explores the neutral network of a given protein starting from
its PDB sequence $\A_0$ and iterating the
following procedure: At time step $t$, (i) The number of viable neighbors of
sequence $\A_t$ is computed; (ii) The sequence $\A_{t+1}$ is extracted at
random among all the viable neighbors of $\A_t$. In this way
we generate a stochastic process along the neutral network which simulate
neutral evolution and looses memory of the initial sequence very fast.

Sequence $\A$ is regarded as viable if both parameters $\a(\A,\C^*)$ and
$-Z'(\A,\C^*)$ are above predetermined thresholds, chosen as 98.5 percent of
the values of those parameters for the sequence in the PDB. This enforces
conservation of the thermodynamic stability and folding capability of the
native structure $\C^*$. We verified that the
observed behavior does not change qualitatively for thresholds between 95\%
and about 100\% of the PDB values.

We impose strict conservation of the cysteine residues in the PDB sequence,
and do not allow any residue to mutate to cysteine, since a mutation changing
the number of cysteine residues would leave the protein with a very reactive
impaired cysteine that would probably affect its functionality. Accordingly,
the total number of neighbors tested is $X_{\rm tot}=18(N-N_{\rm cys})$, where
$N$ is the number of residues and $N_{\rm cys}$ is the number of cysteine
residues in the starting sequence.

The total number of viable point mutations, $X(\A)$, expresses the local
connectivity of the neutral network. We normalize it by the total number
of neighbors, $X_{\rm tot}$, getting the fraction of neutral neighbors,
$x(\A)=X(\A)/X_{\rm tot}\in (0,1]$.

To compute $x(\A)$, we have to evaluate the $\a$ parameter for all sequences
$\A'$ obtained through a point mutation of sequence $\A$.
From Eq.(\ref{alpha}) we note that the $\alpha$ parameter can be obtained
from the configuration with the highest destabilizing power, i.e. the highest
value of the energy gap divided by the structural distance from the native
configuration. These change from sequence to sequence, but it is expected
not to change very much for neighboring sequences.
Thus, in order to speed up the computation of $\a(\A')$, instead of
considering all candidate configurations we consider only the 50
configurations with the highest destabilizing power (i.e. the energy gap
divided by the structural distance from the native configuration)
for sequence $\A$ and compute their mutated destabilizing power
using sequence $\A'$. The $\a$ parameter is then obtained from the
configuration with the highest destabilizing power. This
procedure could slightly overestimate $\alpha(\A')$ since not all
configurations are used, but we have checked
that the error introduced in the $x$ value is in all cases below 0.1\%.

\subsection*{Substitution process}

Given an evolutionary trajectory $\{x_1,x_2,\cdots \}$, the distribution
of the number of substitutions taking place in a time $T$ can be
computed by considering Eq.(\ref{acc}), where $k$, the number of attempted
mutations, is a Poissonian variable of average value $\mu T$.

In order to handle the computation, we divide all values of
$x_i$ in $M$ classes, choosing $X_a$ as representative value of all $x_i$'s
belonging to class $a$.
The number of operations needed to evaluate the substitution probability
increases exponentially with the number of classes $M$. At the same time, the
evaluation becomes more and more accurate as $M$ increases. We chose $M=6$
in our numerical computations as a reasonable compromise between accuracy
and rapidity, checking that larger values of $M$ introduce only small
changes.

\subsection*{Rigidity}
A measure of the conservation profile for a set of evolutionarily related
sequences can be obtained measuring the rigidity of each position $i$,

\be R(i)=\sum_a f_i(a)^2 \, , \label{rigidity}
\ee
where $f_i(a)$, $a=1,\cdots 20$ is the frequency with which amino acid
$a$ is observed at position $i$, normalized so that $\sum_a f_i(a)\equiv 1$.
Deletion of position $i$ in a sequence is regarded formally as a 21st amino
acid. A large rigidity $R(i)$ means that position $i$ is highly conserved.
For unconstrained positions and in absence of deletions, $f_i(a)=1/M$,
where $M$ is the number of amino acids, and $R(i)=1/M$. In general,
rigidities are larger than $1/M$ because of the finite size of the sequence
set and because sequences in the set are correlated due to common
evolutionary origin. Since cysteine residues are strictly conserved, we
always get $R(i)=1$ for them. Thus we omit these residues from the analysis
of conservation.

\subsection*{PFAM and FSSP databases}

We compare the rigidity measured in the set of neutral sequences generated
with the present method with the rigidity obtained from two databases:
the PFAM database (Bateman \ea, 2000)
%\cite{pfam}
and the FSSP database (Holm \& Sander, 1996).
%\cite{dali}
The PFAM database is a collection of families of homologous sequences
obtained by multiple alignment. Since multiple alignment methods work only
for sufficiently high similarity, there are no sequences of low similarity
in this database. The FSSP database is a collection of protein classes
sharing the
same fold (as determined by the program of structural alignment DALI
(Holm \& Sander, 1996)).
%\cite{dali}
Since the structures must be experimentally known, the FSSP
database is usually smaller than the PFAM database. However it includes in
the same class distant homologs whose evolutionary relationship can not be
detected by means of sequence comparison alone. Due to database biases, many
sequences in the PFAM and FSSP databases are highly similar. To reduce this
effect, proteins with similarity higher than a threshold $q_{\thr}=0.85$ to
any other protein have been eliminated.

\section*{Acknowledgments}
UB thanks for interesting discussions Bill Eaton, Carlos Briones,
Nikolai Dokholyan, Leo Mirny, Victor Parro, Burkhard Rost, Alfonso Valencia
and Roeland Van Ham, and the Institut of
Crystallography of the Free University of Berlin for logistic help.
UB is supported by INTA (Spain).
MP is supported by the Max Planck Gesellschaft (Germany).
HER is supported by INFN (Italy).
MV is supported by EMBO and by the Royal Society (UK).

\newpage
\onecolumn

\begin{table}[h]
\begin{tabular}{|l|l|c|c|c|c|c|c|}\hline
\label{table1}
Protein & PDB id. & $N$ & $-Z$ & $\a$ & $\la x\ra$ & $\sigma(x)$ & $\tau$ \\
\hline
rubredoxin (m.)     & 1iro  & 53  & 0.357 & 0.361 & 0.634 & 0.184 & 1.8 \\
rubredoxin (th.)    & 1brf  & 53  & 0.455 & 0.405 & 0.619 & 0.175 & 2.0 \\
cytochrome c        & 451c  & 82  & 0.403 & 0.462 & 0.548 & 0.193 & 2.2 \\
ribonuclease        & 7rsa  & 124 & 0.410 & 0.424 & 0.666 & 0.188 & 2.2 \\
lysozyme            & 3lzt  & 129 & 0.355 & 0.512 & 0.628 & 0.195 & 2.8 \\
myoglobin           & 1a6g  & 151 & 0.458 & 0.575 & 0.599 & 0.191 & 2.4 \\
ubiquitin           & 1u9aA & 160 & 0.402 & 0.568 & 0.631 & 0.195 & 2.4 \\
TIM barrel          & 7timA & 247 & 0.377 & 0.795 & 0.656 & 0.192 & 2.4 \\
\hline
\end{tabular}
\vspace{2truemm}
\caption{Summary of the seven neutral networks studied. For rubredoxin,
m. and th. stand for the mesophylic and thermophylic form respectively.
$x$ indicates the fraction of neutral neighbors and $\tau$ is the
correlation length of $x$ along an evolutionary trajectory obtained from
the stretched exponential decay of the correlation function.}
\end{table}
\vspace{.5cm}

{\bf Abbreviations:} PDB Protein Data Bank,
SNC Structurally Constrained Neutral Model,
FSSP Fold classification based on Structure-Structure alignment of Proteins.


\begin{thebibliography}{20}

\bibitem{Ayala}
Ayala, F.J. (1997)
Vagaries of the molecular clock
{\it Proc. Natl. Acad. Sci. USA \bf 94}: 7776-7783.

\bibitem{altamirano00}
Altamirano, M.M., Blackburn, J.M., Aguayo C. \& Fersht, A.-R. (2000)
Directed evolution of new catalytic activity using the
alpha/beta-barrel scaffold,
{\it Nature \bf 403}, 617-622.

\bibitem{babadje}
Babajide, A., Hofacker, I.L., Sippl, M.J. \& Stadler, P.F. (1997)
Neutral networks in protein space,
{\it Fol. Des. \bf 2}, 261-269.

\bibitem{neutral0}
Bastolla, U. \& Peliti, L. (1991),
Un modele statistique d'evolution avec selection
stabilisante, {\it C. R. Acad. Sci. Paris}, 313, Serie III, 101-105.

\bibitem{opt2}
Bastolla, U., Farwer, J., Knapp, E.W. \& Vendruscolo, M. (2001)
How to guarantee optimal stability for most representative structures in the
protein data bank,
{\it Proteins \bf 44}, 79-96.

\bibitem{UME99}
Bastolla, U., Roman, H.E. \& Vendruscolo, M. (1999)
Neutral evolution of model proteins: Diffusion in sequence space and
overdispersion,
{\it J. Theor. Biol. \bf 200}, 49-64.

\bibitem{opt1}
Bastolla, U., Vendruscolo, M. \& Knapp, E.W. (2000a)
A statistical mechanical method to optimize
energy functions for protein folding,
{\it Proc. Natl. Acad. Sci. USA \bf 97}, 3977-3981.

\bibitem{UME00}
Bastolla, U., Vendruscolo, M. \& Roman, H.E. (2000b)
Structurally constrained protein evolution: Results from a lattice
simulation,
{\it Eur. Phys. J. B \bf 15}, 385-397.

\bibitem{pfam}
Bateman, A., Birney, E., Durbin, R., Eddy, S.-R., Howe, K.-L.
\& Sonnhammer, E.-L.-L. (2000)
The PFAM contribution to the annual NAR database issue,
{\it Nucl. Ac. Res. \bf 28}, 263-266.
Databases available at {\tt http://pfam.wustl.edu/}.

\bibitem{BB}
Bornberg-Bauer, E. (1997)
How are model protein structures distributed in sequence space?,
{\it Biophys. J. \bf 73}, 2393-2403;
Bornberg-Bauer, E. \& Chan, H.S. (1999)
Modeling evolutionary landscapes: Mutational stability, topology, and
superfunnels in sequence space,
{\it Proc. Natl. Acad. Sci. USA \bf 96}, 10689-10694.

\bibitem{Z}
Bowie, J.U., L\"uthy, R. \& Eisenberg, D. (1991)
A method to identify protein sequences that fold into a known 3-dimensional
structure,
{\it Science \bf 253}, 164-170.

\bibitem{britten86}
Britten R.J. (1986)
Rates of DNA sequence evolution differ between taxonomic groups,
{\it Science \bf 231}, 1393-1398.

\bibitem{thirumalai}
Bussemaker, H.J., Thirumalai, D. \& Bhattacharjee, J.K. (1997)
Thermodynamic stability of folded proteins against mutations,
{\it Phys. Rev. Lett. \bf 79}, 3530-3533.

\bibitem{fabio}
Cecconi, F., Micheletti, C., Carloni, P. \& Maritan, A. (2001)
Molecular dynamics studies on HIV-1 protease,
{\it Proteins \bf 43}, 365-372.

\bibitem{Niko}
Dokholyan, N.V. \& Shakhnovich, E.I. (2001)
Understanding hierarchical protein evolution from first principles,
preprint {\tt cond-mat/0104469}.

\bibitem{Fontana}
Fontana, W. \& Schuster, P. (1998)
Continuity in evolution: On the nature of transitions,
{\it Science \bf 280}, 1451-1455.

\bibitem{gerstein98}
Gerstein, M. (1998)
Patterns of protein-fold usage in eight microbial genomes:
A comprehensive structural census,
{\it Proteins \bf 33}, 518-534.

\bibitem{gillespie89}
Gillespie, J.H. (1989)
Lineage effects and the index of dispersion of molecular evolution,
{\it Mol. Biol. Evol. \bf 6}, 636-647.

\bibitem{gillespie91}
Gillespie, J.H. (1991)
The causes of molecular evolution,
Oxford University Press.

\bibitem{Z2}
Goldstein, R.A., Luthey-Schulten Z.A. \& Wolynes, P.G. (1992)
Optimal protein-folding codes from spin-glass theory,
{\it Proc. Natl. Acad. Sci. USA \bf 89}, 4918-4922.

\bibitem{gold1}
Govindarajan, S. \& Goldstein, R.A. (1997)
The foldability landscape of model proteins,
{\it Biopolymers \bf 42}, 427-438.

\bibitem{gold2}
Govindarajan, S. \& Goldstein, R.A. (1998)
On the thermodynamic hypothesis of protein folding,
{\it Procl. Natl. Acad. Sci. USA \bf 95}, 5545-5549.

\bibitem{henikoff97}
Henikoff, S., Greene, E.-A., Pietrokovski, S., Bork, P.,
Attwood, T.-K. \& Hood, L. (1997)
Gene families: The taxonomy of protein paralogs and chimeras,
{\it Science \bf 278}, 609-614.

\bibitem{pdb_select}
Hobohm, U. \& Sander, C. (1994)
Enlarged representative set of protein structure,
{\it Protein Sci. \bf 3}, 522-524.

\bibitem{dali}
Holm, L. \& Sander, C. (1996)
Mapping the protein universe,
{\it Science \bf 273}, 595-602
and {\tt http://www2.ebi.ac.uk/dali/fssp/}.

\bibitem{Huynen}
Huynen, M.A., Stadler, P.F. \& Fontana, W. (1996)
Smoothness within ruggedness: The role of neutrality in adaptation,
{\it Proc. Natl. Acad. Sci. USA \bf 93}, 397-401.

\bibitem{vishveshwara}
Kannan, N. \& Vishveshwara, S. (1999)
Identification of side-chain clusters in protein structures by a graph
spectral method,
{\it J. Mol. Biol. \bf 292}, 441-464;
(2000) Aromatic clusters: A determinant of thermal stability of thermophylic
proteins, {\it Prot. Eng. \bf 13}, 753-761.

\bibitem{keefe01}
Keefe, A.-D. \& Szostak, J.W. (2001)
Functional proteins from a random sequence library,
{\it Nature \bf 410}, 715-718.

\bibitem{kimura68}
Kimura, M. (1968)
Evolutionary rate at the molecular level,
{\it Nature \bf 217}, 624-626.

\bibitem{kimura77}
Kimura, M. (1977)
Preponderance of synonimous changes as evidence for
the neutral theory of molecular evolution.
{\it Nature \bf 267}, 275-6.

\bibitem{king69}
King, J.-L. \& Jukes, T.H. (1969)
Non-Darwinian evolution,
{\it Science \bf 164}, 788-798.

\bibitem{koonin00}
Koonin, E.-V., Aravind, L. \& Kondrashov, A.-S. (2000)
The impact of comparative genomics on our understanding of evolution,
{\it Cell \bf 101}, 573-576.

\bibitem{langley73}
Langley, C.-H. \& Fitch, W.M. (1973)
An estimation of the constancy of the rate of molecular evolution,
{\it J. Mol. Evol. \bf 3}, 161-177

\bibitem{li87}
Li, W.H., Tanimura, M. \& Sharp, P.M. (1987)
An evaluation of the molecular clock hypothesis using mammalian DNA
sequences,
{\it J. Mol. Evol. \bf 25}, 330-342.

\bibitem{MS}
Mirny, L. \& Shakhnovich, E.I. (1996)
How to derive a protein folding potential? A new approach to an old problem,
{\it J. Mol. Biol. \bf 264}, 1164-1179.

\bibitem{ms99}
Mirny, L.A. \& Shakhnovich, E.I. (1999)
Universally conserved positions in protein folds: Reading evolutionary signals
about stability, folding kinetics, and function,
{\it J. Mol. Biol. \bf 291}, 177-196.

\bibitem{ms01}
Mirny, L. \& Shakhnovich, E. (2001)
Evolutionary conservation of the folding nucleus,
{\it J. Mol. Biol. \bf 308}, 123-129.

\bibitem{orencia01}
Orencia, M.-C., Yoon, J.-S., Ness, J.-E., Stemmer, W.-P. \& Stevens, R.-C.
(2001)
Predicting the emergence of antibiotic resistance by directed evolution and
structural analysis,
{\it Nat. Struct. Biol. \bf 8}, 238-242.

\bibitem{ohta76}
Ohta, T. (1976)
Role of very slightly deleterious mutations in molecular evolution
and polymorphism,
{\it Theor. Pop. Biol. \bf 10}, 254-275.

\bibitem{ohta71}
Ohta, T. \& Kimura, M. (1971)
On the constancy of the evolutionary rate of cistrons,
{\it J. Mol. Evol. \bf 1}, 18-25.

\bibitem{ptitsyn98}
Ptitsyn, O.B. (1998)
Protein folding and protein evolution: Common folding nucleus
in different subfamilies of c-type cytochrome?
{\it J. Mol. Biol. \bf 278}, 655-666.

\bibitem{ptitsyn99}
Ptitsyn, O.B. \& Ting, K.H. (1999)
Non-functional conserved residues in globins and their possible
role as a folding nucleus,
{\it J. Mol. Biol. \bf 291}, 671-682.

\bibitem{rost97}
Rost, B. (1997)
Protein structures sustain evolutionary drift,
{\it Fol. Des. \bf 2}, S19-S24.

\bibitem{RNA}
Schuster, P., Fontana, W., Stadler, P.F. \& Hofacker, I.L. (1994)
From sequences to shapes and back -- A case-study in RNA secondary
structures,
{\it Proc. R. Soc. London B \bf 255}, 279-284;

\bibitem{shak96}
Shakhnovich, E., Abkevich, V. \& Ptitsyn, O. (1996)
Conserved residues and the mechanism of protein folding,
{\it Nature \bf 379}, 96-98.

\bibitem{SGinv}
Shakhnovich, E.I. \& Gutin, A.M. (1993)
Engineering of stable and fast-folding sequences of model proteins,
{\it Proc. Natl. Acad. Sci. USA \bf 90}, 7195-7199;
Shakhnovich E.I. (1994)
Proteins with selected sequences fold into unique native conformation,
{\it Phys. Rev. Lett. \bf 72}, 3907-3910.

\bibitem{silverman01}
Silverman, J.A., Balakrishan, R. \& Harbury, P.B. (2001)
Reverse engineering the $(\beta/\a)_8$ barrel fold,
{\it Proc. Natl. Acad. Sci. USA \bf 98}, 3092-3097.

\bibitem{takahata87}
Takahata, N. (1987)
On the overdispersed molecular clock,
{\it Genetics \bf 116}, 169-179.

\bibitem{tiana}
Tiana, G., Broglia, R.A., Roman, H.E., Vigezzi, E. \& Shakhnovich, E.I. (1998)
Folding and misfolding of designed proteinlike chains with mutations,
{\it J. Chem. Phys. \bf 108}, 757-761.

\bibitem{newitem}
Thornton, J.M., Orengo, C.A., Todd, A.E. \& Pearl, F.M.G. (1999)
Protein folds, functions and evolution
{\it J. Mol. Biol. \bf 293}, 333-342.

\bibitem{voigt01}
Voigt, C.A., Mayo, S.L., Arnold, F.H. \& Wang, Z.G. (2001)
Computational method to reduce the search space for directed protein
evolution,
{\it Proc. Natl. Acad. Sci. USA \bf 98}, 3778-3783.

\bibitem{ZP}
Zuckerkandl, E. \& Pauling, L. (1962), in {\it Horizons in Biochemistry},
eds. M. Kasha and B. Pullman (Academic Press, New York).


\end{thebibliography}
\end{document}